# Topological Effect of Surface Plasmon Excitation in Gapped Isotropic Topological Insulator Nanowires

Mingda Li[1]*, Wenping Cui[2,6], Lijun Wu[3], Qingping Meng[3], Yimei Zhu[3], Yong Zhang[4], Weishu Liu[5], Zhifeng Ren[5]

[1]*Department of Nuclear Science and Engineering, Massachusetts Institute of Technology, Cambridge, MA 02139, USA*

[2]*Department of Physics, University of Bonn, 53113 Bonn, Germany*

[3]*Center for Functional Nanomaterials, Brookhaven National Lab, Upton, NY11973, USA*

[4]*Center for Materials Science and Engineering, Massachusetts Institute of Technology, Cambridge, MA 02139, USA*

[5]*Department of Physics, Boston College, Chestnut Hill, MA 02467, USA*

[6]*Institute for Theoretical Physics, University of Cologne, Cologne 50937 , Germany*

*\* Author to whom correspondence should be addressed. Email: mingda@mit.edu*

## Abstract

We present a theoretical investigation of the surface plasmon (SP) at the interface between topologically non-trivial cylindrical core and topological-trivial surrounding material, from the axion electrodynamics and modified constitutive relations. We find that the topological effect always leads to a red-shift of SP energy, while the energy red-shift decreases monotonically as core diameter decreases. A qualitative picture based on classical perturbation theory is given to explain these phenomena, from which we also infer that in order to enhance the shift, the difference between the inverse of dielectric constants of two materials shall be increased. We also find that the surrounding magnetic environment suppresses the topological effect. All these features can be well described by a simple ansatz surface wave, which is in good agreement with full electromagnetic eigenmodes. In addition, bulk plasmon energy at $\omega_P = 17.5 \pm 0.2 eV$ for semiconducting $Bi_2Se_3$ nanoparticle is observed from high-resolution Electron Energy Loss Spectrum Image measurements.

## I. Introduction

Surface plasmon (SP) is an electronic collective excitation localized at the interface between two materials. Such a collective oscillation is induced by long-range Coulomb interactions between near-boundary electrons [2, 3]. Classically, SP is regarded as a propagating or standing wave at the interface between two materials with opposite sign of dielectric functions.

SP has been extensively studied in metallic nanoparticles [4-8] and found wide applications in nanophotonics and nanoplasmonics [4, 7, 9, 10]. Recently, SP generated by swift electrons, in metallic nanoparticles with various shapes, such as spheres [6, 11, 12], cubes [12] and rods [13-15], have attracted much attention, since nanostructures with different geometries show distinct resonant SP modes. Moreover, electron-excited SP reveals dark SP modes which cannot be resolved by previous optical-excited experiments [13].

Topological insulator (TI), on the other hand, is one kind of topologically-nontrivial materials, with insulating state in the bulk and chiral metallic states at the surface [16, 17]. Therefore, SP excitation is expected due to the metallic nature at the TI surface.

Despite the blooming studies of SP at the interface between two topologically-trivial materials, with various geometries, there are still very few reports considering the SP at the interface between materials with different topological orders. Efimkin *et al* [18] develops a quantum field theory of spin-plasmon on the surface of TI based on random phase approximation, in which case the gap of the surface states is not opened while the electron spin and momentum are always locked. On the other hand, Karch *et al* [19, 20] develops a classical theory of the SP on TI slab and thin film within the frame of axion electrodynamics, where the electromagnetic Lagrangian adds another axion term $\theta \vec{E} \cdot \vec{B}$ [21], which is valid when the gap is opened and leads to an exotic magnetoelectric effect. In this classical picture, the spin of electrons is not taken into account. Despite that these two theories start from gapless or gapped state, within the frame of either quantum field or classical theory, they both focus on ideal slab-geometry only. To the best of our knowledge, SP in TI of with other geometries has not been reported.

In this article, we adopt the latter axion electrodynamics approach, but apply it into a more realistic cylindrical geometry. We consider a TI nanowire with isotropic dielectric function, as an infinite long cylinder, immersed in dielectric materials with different magnetic permeabilities. We calculate the shift of SP energy from the SP dispersion relation due to topological effect (i.e. the axion term), and find that topological effect always lowers the SP energy by an order of $\sim \alpha^2$, where   is the fine structure constant. We also find that when the radius of the nanowire becomes smaller, the topological modification of SP energy is suppressed. Both phenomena are explained within a qualitative approach of perturbation theory for Maxwell's equations with shifting boundaries [22]. By stretching an infinite cylindrical nanowire uniformly, it results in another infinite cylinder with smaller radius. This means that different core radii are equivalent to stretched boundary.

In order to dig out the significance of the topological axion-term, we use a simple surface wave form as an ansatz, which propagates along the cylinder and is evanescent away from the interface, and compare the result with the rigorous electromagnetic eigenmodes in cylindrical geometry. This ansatz correctly grasps all the features of the topological effect to SP dispersion.

What's more, since the bulk plasmon (BP) energy $\omega_P$ is the only parameter entering into the dispersion relation calculation, we measure the BP energy for nanoparticle $Bi_2Se_3$, from high-resolution electron energy loss spectroscopy (EELS) studies. The BP energy is then obtained by averaging the value over all spatial points in a 2D EELS mapping. The value $\omega_P = 17.5 \pm 0.2 eV$ is a little greater than the value of $\omega_P = 16.8 eV$ for single crystalline $Bi_2Se_3$ [23] due to a larger carrier density in the nanoparticles.

## II. Theory

### A. Topological dispersion relation for Cylindrical Surface Wave

For the interface between topological insulator and ordinary material, the Maxwell's equations themselves are not changed, yet the constitutive relations are modified to couple the responses of E and B field together, as long as the Dirac surface state is gapped [20, 21]. In SI units, the modified constitutive relations to linear order can be written as:

$$\begin{cases} \vec{D} = \varepsilon \vec{E} - \varepsilon_0 \alpha \dfrac{\theta}{\pi} c\vec{B} \\ c\vec{H} = \dfrac{c\vec{B}}{\mu} + \alpha \dfrac{\theta}{\pi} \dfrac{\vec{E}}{\mu_0} \end{cases} \quad (1)$$

where θ is the pseudoscalar part of the E-B coupling [24], $\theta = \pi$ for TI, while $\theta = 0$ for topologically-trivial materials. $\alpha = \dfrac{1}{137.036}$ is the fine structure constant, c is the light speed in vacuum, $\varepsilon$ and $\mu$ are dielectric constant and magnetic permeability in that material, respectively.

In order to solve the Maxwell's equations in cylindrical geometry with coupled constitutive relations eq. (1), we start from an ansatz of a surface wave in cylindrical

coordinate, which propagates along the cylinder z-axis and decays exponentially away from the interface:

$$\begin{cases} \vec{E}_i = \vec{E}_i(r,\varphi) e^{-k_i|r-d|} e^{i(qz-\omega t)} \\ \vec{B}_i = \vec{B}_i(r,\varphi) e^{-k_i|r-d|} e^{i(qz-\omega t)} \end{cases} \quad (2)$$

where $i=1,2$, with 1 denotes the outer material while 2 denotes the cylindrical core, $d$ is the radius of the nanowire, $q$ is the wavenumber along the propagating z direction, $k_i$ $i=1,2$ are the decay constants with $k_i>0$ hold.

Both E and B fields have component in all (r, , z) directions, but each component is a function of r and only due to translational symmetry along z-direction.

Despite the ansatz is not the eigenmode in cylindrical geometry, it is still very worthwhile to be studied, with the following reasons:

1) The cylindrical eigenmodes have exponential-like decay behavior away from the interface originated from Modified Bessel functions, while the ansatz has similar behavior but uses a simpler exponential-decay function.
2) Based on Drude model of dielectric constant, this ansatz shows two dispersions, which can be identified as N=0 mode and superposition of N>0 modes (compare Fig. 1(b) with Fig. 4, where N is the eigen-number of the electromagnetic eigenmodes.
3) It can be reduced to the flat-slab case directly by simply taking $d \to \infty$ in the diameter-dependent dispersion relation, while the eigen solutions cannot.

Substituting the ansatz form back to the Maxwell's equation in cylindrical coordinate, we obtain

$$\left(\frac{\alpha\theta}{\pi}\right)^2 \frac{1}{c^2\mu_0^2} = -\left(\frac{1/d-k_1}{\mu_1} - \frac{1/d+k_2}{\mu_2}\right)\left(\frac{\varepsilon_1}{1/d-k_1} - \frac{\varepsilon_2}{1/d+k_2}\right) \quad (3)$$

where $\theta \equiv \theta_2 - \theta_1 = \pi$ is the difference of topological number.

By taking $d \to \infty$, eq. (3) directly reduces to the topological SP dispersion for a flat slab $\pi^2\mu_0\left(\frac{k_1}{\mu_1}+\frac{k_2}{\mu_2}\right)\left(\frac{\varepsilon_1}{k_1}+\frac{\varepsilon_2}{k_2}\right) = -\varepsilon_0\alpha^2\theta^2$, which is reported in [20].

By setting $\theta = 0$, this dispersion relation further reduces to the traditional condition of transverse magnetic (TM) mode-SP $\frac{\varepsilon_1}{k_1}+\frac{\varepsilon_2}{k_2}=0$ and transverse electric (TE) mode-SP $\frac{k_1}{\mu_1}+\frac{k_2}{\mu_2}=0$, directly [3].

## B. Topological-SP dispersion relation with ideal Drude model

Since $k_i>0$, $i=1,2$, the TE mode requires $\mu_1$ and $\mu_2$ have opposite sign, which is difficult to achieve. On the other hand, the opposite sign condition of $\varepsilon_1$ and $\varepsilon_2$ for TM mode can be achieved at metal-dielectric interface as long as the frequency is lower than bulk-plasmon frequency $\omega_P$, in which case the dielectric constant for the metal-part can be described by Drude model [25]. In the situation related to TI, the SP is allowed at the metal-TI interface when the TI part is insulating, and at vacuum-TI interface when the TI has finite bulk carrier density [20]. In the latter case, the semiconducting bulk TI can be described through modified Drude model [26], for simplicity the frequency-dependent Drude ideal model is applied:

$$\frac{\varepsilon_2}{\varepsilon_0} = 1 - \frac{\omega_P^2}{\omega^2} \tag{4}$$

The tolerance of finite bulk carrier density of TI to observe SP has advantage compared to transport measurement, since in a transport measurement, the bulk carrier density has be kept as low as possible in order to make the surface transport dominate [27, 28].

Redefine all quantities in dimensionless unit, $q \to \frac{qc}{\omega_P}$, $\omega \to \frac{\omega}{\omega_P}$, $a \to \frac{c}{2d\omega_P}$, the SP dispersion relation eq. (3) with Drude model applied to core region can be written as

$$\left(\frac{\alpha\theta}{\pi}\right)^2 = -\left(\sqrt{a^2+(q^2-\omega^2+1)} + \sqrt{a^2+(q^2-\omega^2)}\right)\left[\frac{1}{\sqrt{a^2+(q^2-\omega^2)}-a} + \frac{\left(1-\frac{1}{\omega^2}\right)}{\sqrt{a^2+(q^2-\omega^2+1)}+a}\right] \tag{5}$$

When the topological term $\frac{\alpha\theta}{\pi}$ and the curvature term $a$ are both set to zero, eq. (5) is reduced to the normal dispersion $q = \omega\sqrt{\frac{\omega^2-1}{2\omega^2-1}}$ for Drude semi-infinite metal in vacuum [3].

## C. Eigenmode in Cylindrical Geometry

The electromagnetic eigenmodes in cylindrical geometry can be written as [1, 29]:

$$E_{ir} = \left[\frac{iq_z}{k_i}a_{ni}f'_{ni}(k_ir) - \frac{\mu_i\omega n}{k_i^2 r}b_{ni}f_{ni}(k_ir)\right]e^{i(n\theta+q_zz-\omega t)}, E_{iz} = \left[f_{ni}(k_ir)a_{ni}\right]e^{i(n\theta+q_zz-\omega t)}$$

$$E_{i\theta} = -\left[\frac{nq_z}{k_1^2 r}a_{ni}f_{ni}(k_i r) + \frac{i\mu_i\omega}{k_i}b_{ni}f'_{ni}(k_i r)\right]e^{i(n\theta+q_z z-\omega t)} \quad (6)$$

$$B_{ir} = \left[\frac{\varepsilon_i\mu_i\omega n}{k_i^2 c^2 r}a_{ni}f_{ni}(k_i r) + \frac{i\mu_i q_z}{k_i}b_{ni}f'_{ni}(k_i r)\right]e^{i(n\theta+q_z z-\omega t)}, B_{iz} = \mu_i\left[f_i(k_i r)b_{ni}\right]e^{i(n\theta+q_z z-\omega t)}$$

$$B_{i\theta} = \left[\frac{i\varepsilon_i\mu_i\omega}{k_i c^2}a_{ni}f'_{ni}(k_i r) - \frac{n\mu_i q_z}{k_i^2 r}b_{ni}f_{ni}(k_i r)\right]e^{i(n\theta+q_z z-\omega t)}$$

where i=1,2 denotes the outer environment and inner core, respectively, $a_n$ and $b_n$ are complex coefficients, $f_{n1} \equiv K_n$, $f_{n2} \equiv I_n$ are modified Bessel functions, $k_i^2 = \varepsilon_i\mu_i\frac{\omega^2}{c^2} - q_z^2$.
For real SP modes, $k_i$ are purely imaginary since they lie below the light line in the dispersion relations. The arguments of the modified Bessel functions are taken as absolute value of $k_i r$, while the differentiation of Bessel functions are respect to the imaginary arguement $k_i r$ itself.

By noticing that even in the topological eignemode case, there are still only four independent boundary conditions, we obtain the dispersion relation

$$\frac{c^2 q_z^2 n^2}{d^2}\left(\frac{1}{k_1^2} - \frac{1}{k_2^2}\right)^2 - \omega^2\left[\frac{I'_n}{I_n}\frac{\varepsilon_2}{k_2} - \frac{K'_n}{K_n}\frac{\varepsilon_1}{k_1}\right]\left[\frac{I'_n}{I_n}\frac{\mu_2}{k_2} - \frac{K'_n}{K_n}\frac{\mu_1}{k_1}\right] + \alpha^2\frac{I'_n}{I_n}\frac{K'_n}{K_n}\frac{\mu_2\mu_1}{k_1 k_2}\omega^2 = 0 \quad (7)$$

Non-trivial solutions of the coefficients can be found if the determinant of the system of linear homogenerous equations is zero. This leads to the following dispersion relation:

Here the arguments of modified Bessel functions are omitted. By setting the last term, which is 2nd-order to    and contains the topological effect, to be zero, eq. (7) reduces to the normal SP dispersion relation in cylindrical geometry, immediately [30]. We also see that in eq. (7), the 1st term and 2nd term in eq. (7) are  -independent and 1st-order of  , respectively, thus in order to make the topological-induced energy shift to plasmon dispersion larger, the product $\mu_2\mu_1$ should be made as large as possible. This may be achieved through magnetic TI [31-34]- ferromagnetic metal interface. Unfortunately, current magnetic TI are produced from magnetic doping so the magnetic suspectibility is still small compared to long-range order ferromagnetism.

## D. Shifted Boundary Condition and Perturbation of Maxwell's Equation

In order to see how the size and topological effect modify the dispersion relation, we apply the analysis of Perturbation theory of Maxwell's equations of shifted boundary [22]. Since strentching an infinite-long cylinder along the symmetry axis results in another infinite-long cylinder with smaller radius, perturbation theory of boundary shifting should be appropriate to describe wires with different radii.

A perturbation of the boundary eventually leads to a perturbation of dielectric constant $\Delta\varepsilon$. The first order correction to the unperturbed eigenfrequency $\omega^{(0)}$ can be written as [22]:

$$\omega^{(1)} = -\frac{\omega^{(0)}}{2}\frac{\left\langle E^{(0)}\left|\Delta\varepsilon\right|E^{(0)}\right\rangle}{\left\langle E^{(0)}\left|\varepsilon\right|E^{(0)}\right\rangle} \tag{8}$$

In the situation of perturbation with shited-boundary of amount $\Delta h$ normal to the shifted surface, the numerator can be further written as

$$\left\langle E^{(0)}\left|\Delta\varepsilon\right|E^{(0)}\right\rangle = \int dA \Delta h \left(\Delta\varepsilon_{12}|E^{(0)}_{//}|^2 - \Delta\varepsilon_{12}^{-1}|D^{(0)}_{\perp}|^2\right) \tag{9}$$

where $\Delta h$ is the shift of the boundary in normal direction, in the cylindrical geometry, $\Delta h = d_{new} - d_{old}$ is the difference of radii after and before the boundary is changed. $E_{//}$ is the component parallel to the shifted boundary, $D_{\perp}$ is the component normal to the shifted boundary, $\Delta\varepsilon_{12} \equiv \varepsilon_1 - \varepsilon_2$ and $\Delta\varepsilon_{12}^{-1} \equiv \varepsilon_1^{-1} - \varepsilon_2^{-1}$.

From eq. (8) and eq. (9), the effect of the variation of cylinder radii together with and topological effect are clear. By substituting each components of electric field in eq. (2) into eq. (9), we can see that when the diameter becomes smaller, $\left\langle E^{(0)}\left|\Delta\varepsilon\right|E^{(0)}\right\rangle > 0$. Therefore, from eq. (8), the 1st-order energy is always negative, which indicates a decrease of energy as the diameter becomes smaller. Moreover, substituting 1st equation of eq. (1) which contains the topological term back to eq. (9), it can be seen directly that the topological term will further lower the energy by an order of $\alpha^2$ compared with the non-topological counterparts. Hence, qualitatively, it is reasonable to conclude that the topological term reduces the energy by an order of $\alpha^2$, which is indeed the case (See in Fig. 2).

Despite its sucess in predicting the trend of the size and topological effect to SP energy, we don't apply it directly to acquire the dispersion since it is only valid for infinitesimal change of boundary, but only for a qualitative picture of energy shift. In order to study cylinder structure with a range of radii, this method would be too costly with an increasing error. Instead, we solve the dispersion relations directly at various radii, and obtained reuslts consistent with this picture.

## III. Results and Discussions

When setting the topological term of eq. (5) to zero, the resulted non-topological SP dispersion can be simplified to an algebaric equation:

$$\omega^4 - \left(4q^2 + 2\right)\omega^3 + \left(4q^4 + 1 + 4a^2q^2 + 6q^2\right)\omega^2 - \left(4a^2q^2 + 4q^4 + 2q^2\right)\omega + q^4 = 0 \quad (10)$$

The dispersion relations based on eq. (10) are plotted in Fig. 1 with three different radii, including the flat slab case where $a=0$ ($d \to \infty$). The well-known BP v.s. SP dispersion duet for the slab are plotted in purple dot-dashed line in Fig.1(a), from which we can see that SP frequency decreases monotonically as diameter decreases. This trend is consistent with the perturbation theory analysis in II.D. On the other hand, the BP mode is not suitable for the perturbation analysis, in that the BP modes with finite life time will break the pre-requisite for perturbation analysis, that eigen problem needs to be semidefinite with real values [22]. The bulk-plasmon decay constant (inverse of lifetime) is shown in Fig. 2(a), which corresponds to radiating plasmon.

Besides the bulk and surface plasmon, an additional non-dispersive plasmon mode is developed (Fig. 1(b) upper curves). Actually, these modes are recognized as SP eigenmode (compare with Fig. 4 and Fig. S2) with the eigen-number N>0 [30]. The reason that both N=0 SP mode and N>0 modes exist in eq. (10) is a result of the form of ansatz eq. (2), which can be regarded as a supposition of eigenmodes.

Since we are interested in the topological effect to the real SP mode dispersion relation, only the real SP modes are discussed. Fig.2 (b) illustrates that how the topological effect modifies the dispersion relation eq. (10), using Levenberg-Marquardt algorithm. Where non-topological $(q, \omega_0)$ are used as initial value to optimize for topological-modified dispersion $(q, \omega_{topo})$.

Comparing Fig. 1 (a, b) with Fig. 2 (b), we see that the energy shift $(q, \omega_{topo} - \omega_0)$ has similar shape of the dispersion $(q, \omega_0)$ itself. This can also be explained by perturbation theory. The 1$^{st}$-order perturbed plasmon energy $\omega^{(1)}$ (eq. (8)) has nearly-proportional dependence with un-perturbed $\omega^{(0)}$. At certain $q$ value, when the core radius is smaller (perturbed), the perturbed SP energy $\omega^{(1)}$ follows the same trend with $\omega^{(0)}$. Furthermore, $\omega^{(1)}$ is also perturbed by the topological term (eqs. (1) & (9) ). Therefore, the red-shift of the dispersions show similar behavior to the original dispersions itself.

In order to see how magnetism of the surrounding environment modifies the topological SP dispersion, the SP dispersion and energy shifts as a function of radii and magnetic permeability are plotted in Fig. 3. Despite the fact that the gapped TI surface state is prerequisite to the existence of classical SP mode, large magnetization of outer material, which leads to a larger surface gap, does not help to enhance the topological induced SP energy shift.

Eigen-equations eq. (7) are solved using Newton method. The dispersion relations with N=0,1 are plotted in Fig. 4. Several other eigenmodes are shown in supplymentary material S4. The N=0 mode in Fig. 4 (a) are identical to the lower SP mode in Fig. 1, which is expected, and has a limit behavior of $\omega \to 0$ as $q$ goes to zero. On the other hand, the N>0 modes intersect with the light line at finite , which are also expected from Fig. 1 (b) and [1]. When the radius goes larger, different eigenmodes tend to shrink into one principal curve (Fig. S2 (a)).

The perturbation theory of Maxwell's equation are not applicable to the eigenmodes N>0, since the numerator $\langle E^{(0)} | \Delta \varepsilon | E^{(0)} \rangle$ is not guarenteed to be greater than zero. When N>0, different modes are similar to each other, forming non-dispersive eigenmodes on the right of the light line. This non-dispersive modes are already shown in Fig. 1(b), which is obtained from simple ansatz surface wave other than solving the eigenequations directly; indeed, the exponential-decay behavior of the ansatz wave can then be regarded as a superposition of N=0 eigenmode and higher-N non-dispersive eigenmodes.

The energy shift d=5 nm, and for N=0 eigenmode are shown in Fig. 5. The red-sfhit for N=0 coincides with the exponential-decay ansatz in Fig. 2 (b). In any case, the topological effect red-shifts the energy, and the SP energy shift also shows a similar behavior with the dispersion relation itself. This indicates the the topological behavior for the surface wave of the cylinder is correctly grasped within the frame of the simple exponential-decay ansatz, without the need to solve the complicated eigenmodes each step.

### IV. Determination of BP energy $\omega_P$

Since the BP energy $\omega_P$ is the only external parameter entering into the dispersion relation from Drude model, it is worthwhile to give a reasonable value in order to convert the dimensionless dispersion relation into actual dimension. A BP energy of $\omega_P = 16.8 eV$ of single crystalline strong topological insulator $Bi_2Se_3$ has been reported [23]. Since a finite bulk carrier density is preferred for SP excitation, semiconducting $Bi_2Se_3$ samples are prepared from 20h mechanical alloying and followed by hot-pressing at 500°C.

The high-res EELS is done by Jeol ARM 200F at Center for Functional Nanomaterials in Brookhaven National Lab. TEM image and illustration of EELS line-scan spectrum are shown in Fig. 6 (a) and (b) respectively. As the electron beams scans from the vacuum to deep in the bulk region. This peak can be easily recognized as BP peak, since the Bi *5d* core losses are already well above 20eV [23]. The impurity contribution can be excluded from EDS measurement (supplementary S3), while which indicates an in-significant defect contribution to plasmon excitation. Despite of the observation of BP peak, it is still difficult to resolve other features due to limited resolution and low electron dose.

The height of peak increases with relatively fixed energy position, see in Fig. 6 (d). This is reasonable since thicker inside the sample region, the excitation population is increased, causing an enhancement of peak amplitude. As long as the BP is developed with fixed carrier density, the BP energy should not depend on the position. On the other hand, the peak width is increased near the particle edge, as shown in (e). The peak width is inverse to the life-time of plasmon excitation, therefore, the decrease of BP lifetime near the edgeis likely to be caused by enhanced surface-scattering. In any case, the BP energy can be extracted from the fitting of the BP peak.

Two-dimensional scan of EELS spectrum image is also acquired to extract the BP energy. A 2D area is slected (Fig. 7 (a)), within which the electron beam scans all the position with a 0.7nm electron-beam step size. After fitting the developed BP peaks in the entire region, the BP energies and amplitudes are extracted. The amplitudes are expected to be higher deeper inside the bulk region, while the energy keeps almost constant. Averaging all the points of the BP energy value, we obtain $\omega_P = 17.5 \pm 0.2 eV$ in this particlular local region.

## V. Conclusions

We present the calculation of the topological effect to the dispersion relations of surface plasmon excitation in cylindrical geometry, based on axion electrodynamics and isotropic Drude model. This is applicable to isotropic topological insulator when the surface state is gapped. This condition can be easily achieved, since the propagating surface wave always has a magnetic field component normal to the interface and opens up a gap [35, 36]. Actually, in cylindrical geometry, purely TE or TM wave cannot even propagate [30, 37]. That's why there is always the topological term in the constitutive relation eq. (1): in the classical phenomelological picture, as long as there is surface wave propagating, it will open up a gap and make eq.(1) valid, and a gapless surface wave seems non-existing.

Since the constitute relation eq. (1) is to linear order, there seems to be higher-order contribution to the dispersion relation. However,, the non-linear effect would alter the SP dipserion relation, especially in high-q region, it is unlikely to have contribution to the energy shift. This is because the non-topological dispersion will also be altered correspondingly by non-linear constitutive relation, making the non-topological contribution to dispersion cancel with each other- only the topological term will remain to cause the energy-shift. Furthermore, the energy red-shift is also guarenteed from the analysis of pertrubation theory. Thus, the feature of energy red-shift is preserved even taking non-linear perturbation into account.

In order to see the topological modified SP excitation, the magnetism of the surrounding material is not suffcient, despite intuitively it is helpful to futher break the interface time-reversal symmetry. The prediction from eq. (10) is that the product of magnetic permeability $\mu_1\mu_2$ should be kept as large as possible to make the topological term large enough. However, for the current magnetic TI, such as Cr-doped $Bi_2Se_3$ [38], the ferromagnetism is localized on the small amount of dopant sites, thus the overall magneitc permeability is still too low. If a real ferromagnetic TI is produced, an energy shift should be seen from the SP excitation below and above the ferromagneitc-paramagnetic transition. On the other hand, from eq. (9), in order to make the $D_\perp^{(0)}$ term, which includes the topological term, large enough, the difference between inverse dielectric constant $\Delta\varepsilon_{12}^{-1} = \varepsilon_1^{-1} - \varepsilon_2^{-1}$ should be as made large as possible. However, within the Drude model, $\Delta\varepsilon_{12}^{-1} = \frac{\omega_P^2}{\omega_P^2 - \omega^2}$, while the SP energy $\omega < \omega_P/\sqrt{2}$ in any case. This also becomes a limitation to make the topological-induced SP energy shift observable. On the other hand, if the term $\mu_1\mu_2$ or $\Delta\varepsilon_{12}^{-1}$ can be manipulated much greater, the topological-induced shift in SP energy is õamplifiedö and becomes observable.

Other than the 2nd-order of  in SP energy, eq. (1) indicates that the EM field component change is only 1st-order of . Therefore, this electromagneto effect can be seen from the Faraday and Kerr effect [39] much easily.

## Acknowledges

Author Mingda Li would like to thank Prof. Ju Li for his generous support in this study.

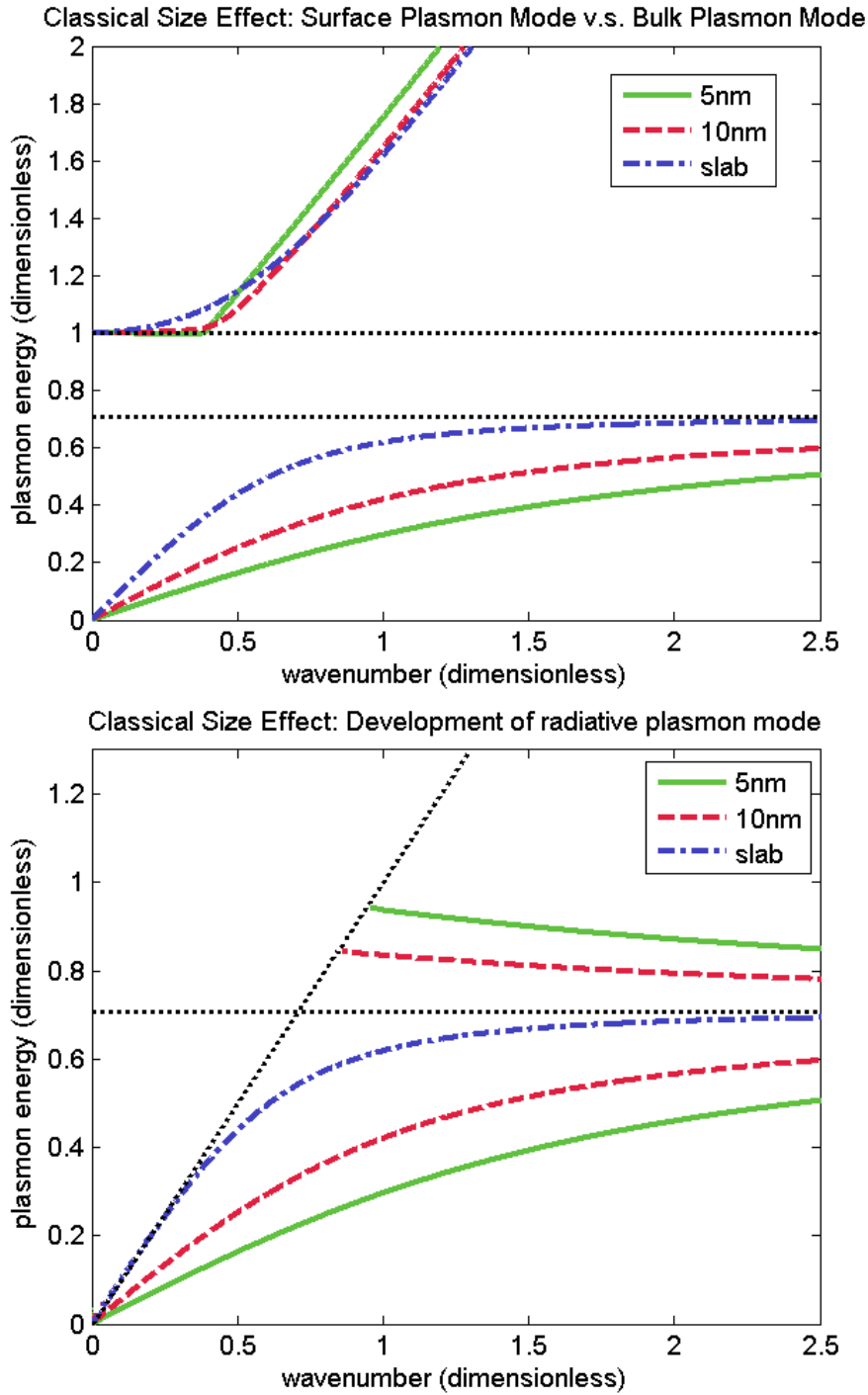

Fig. 1 The plasmon dispersion relations at radius $d$=5nm, 10nm and infinite slab. In (a), the upper curves corresponds to the bulk plasmon dispersion, while the lower curves are surface plasmon modes. In (b), at finite radius, additional non-dispersive modes are also developed, which is recognized as supposition of N>0 SP eigenmodes. For an infinite slab (purple dot-dashed line), all modes are degenerate.

*Fig. 2*

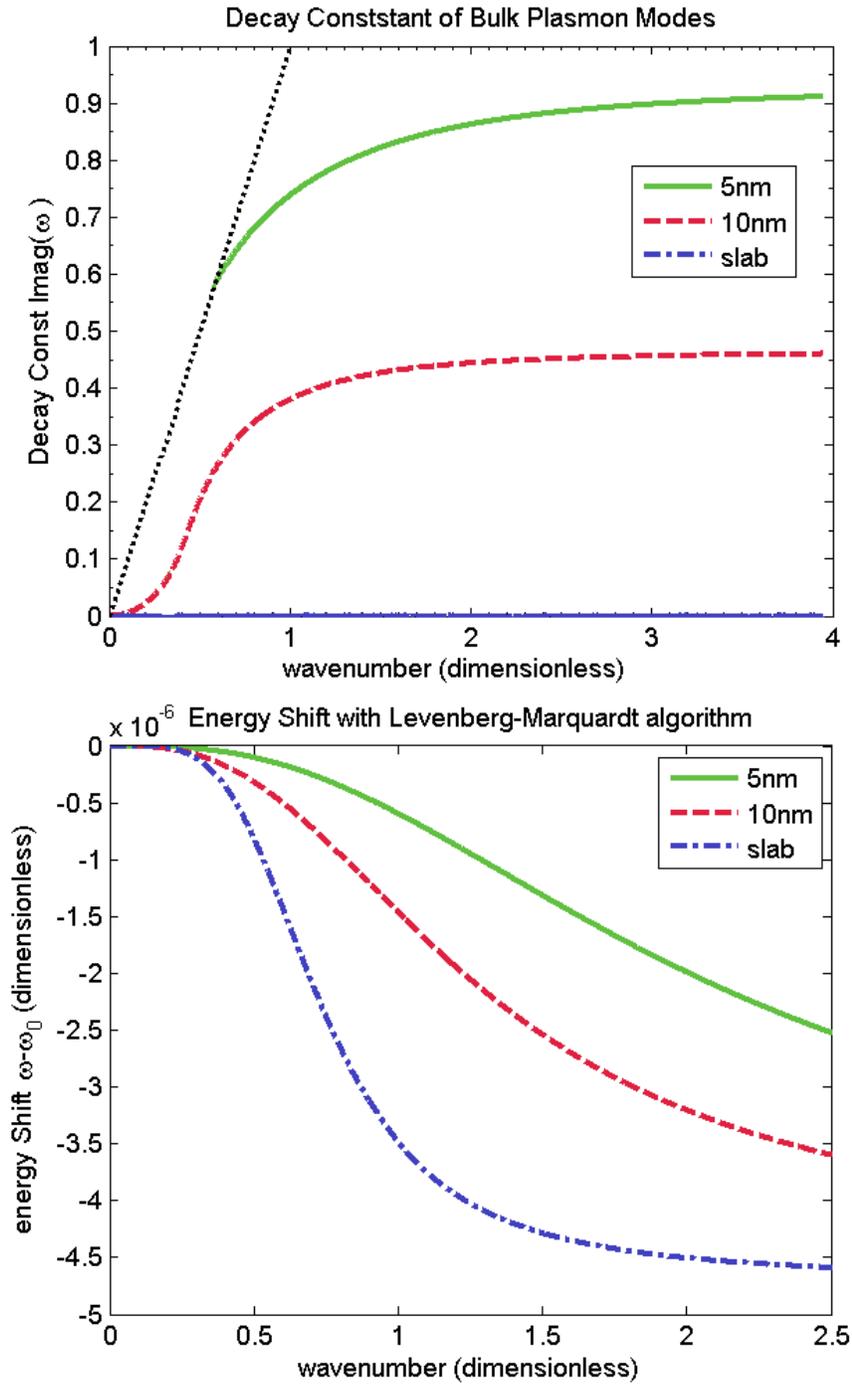

Fig. 2 (a) The decay constant, i.e. the imaginary part of the plasmon energy; finite radius always leads to a decay for BP mode (b) the topological-induced red-shift to plasmon dispersion using Levenberg-Marquardt algorithms.

*Fig.3*

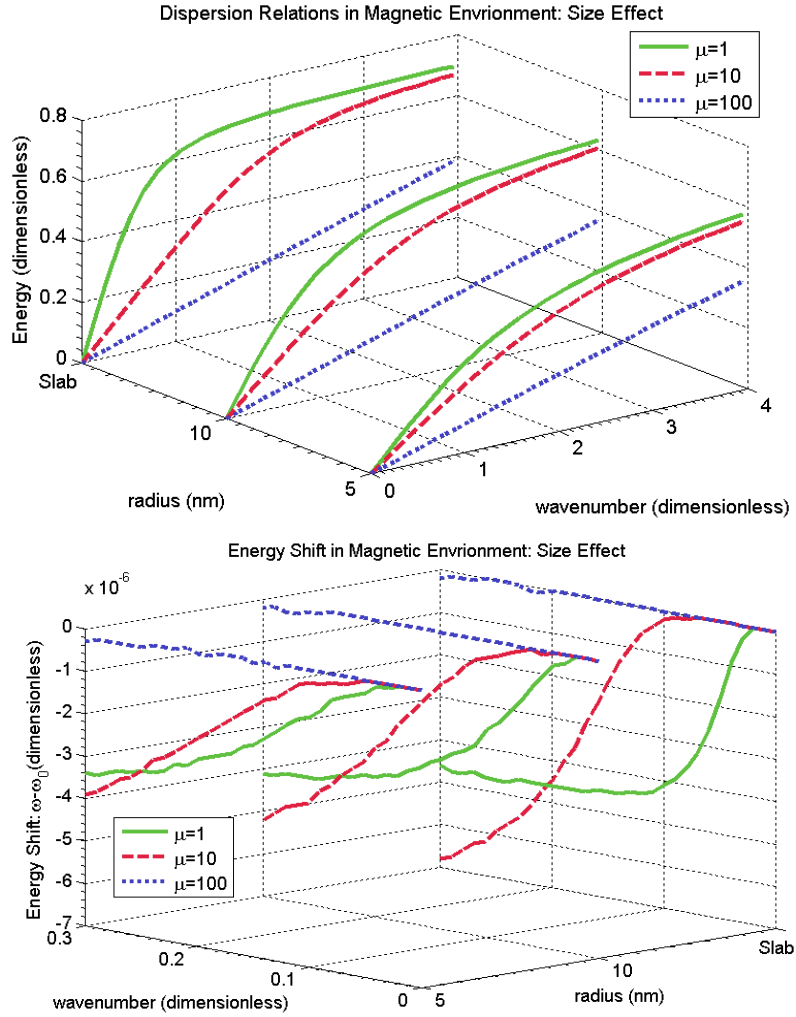

Fig. 3 The surface plasmon dispersion relations (a) and topological-induced energy shifts (b) at magnetic permeability =1,10 and 100. We see that as the magnetic permeability of the outer material goes larger, the SP energy decreases, while the topological-induced change to SP energy tends to vanish.

*Fig. 4*

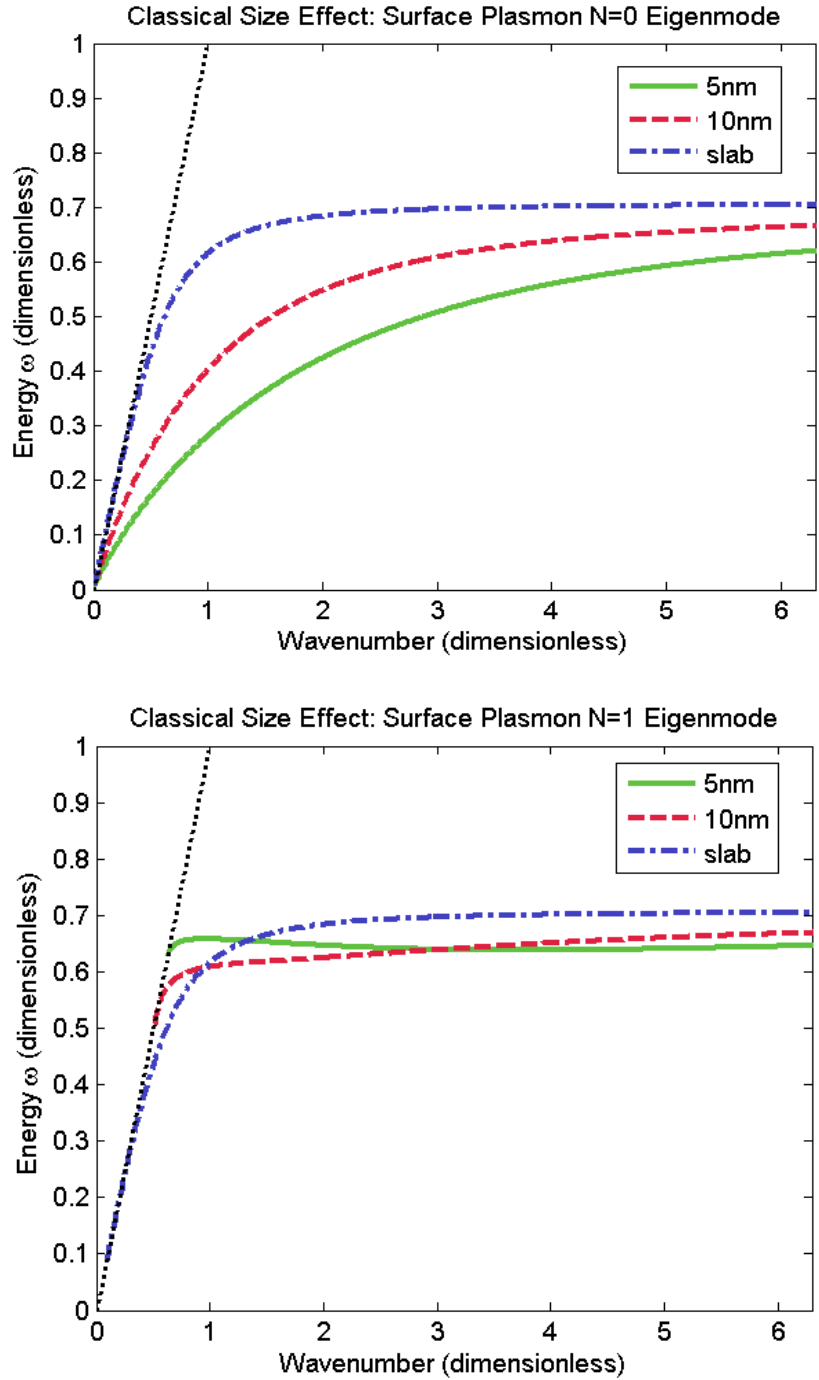

Fig. 4. SP eigenmodes at N=0,1 at various radii. The N=0 modes (a) are similar to the surface plasmon modes solved from the ansatz, and when q goes to zero,  goes to zero accordingly. On the other hand, N=1 modes (b) intersect the light line (denoted in black line) at finite  . When the radius becomes larger, different eigenmodes tend to collapse into one curve, and intersect the light line with smaller energy.

*Fig. 5*

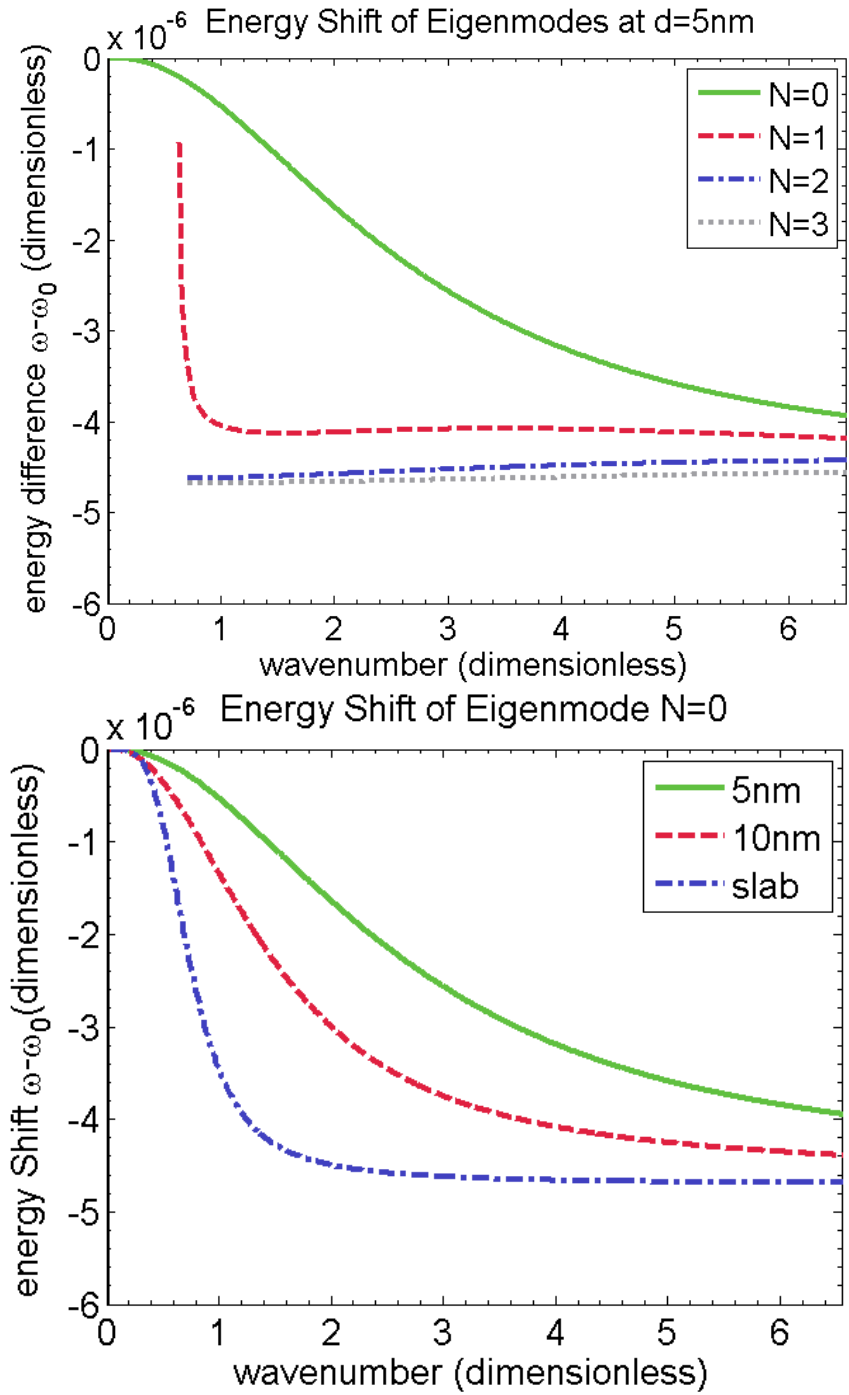

Fig. 5 Dispersion relations at radii *d*=5nm and 10nm, at various eigen number. When the radius goes larger, different modes becomes closer to each other, and finally collapse into one slab-dispersion curve, as shown in supplementary material Fig. S2.

*Fig. 6*

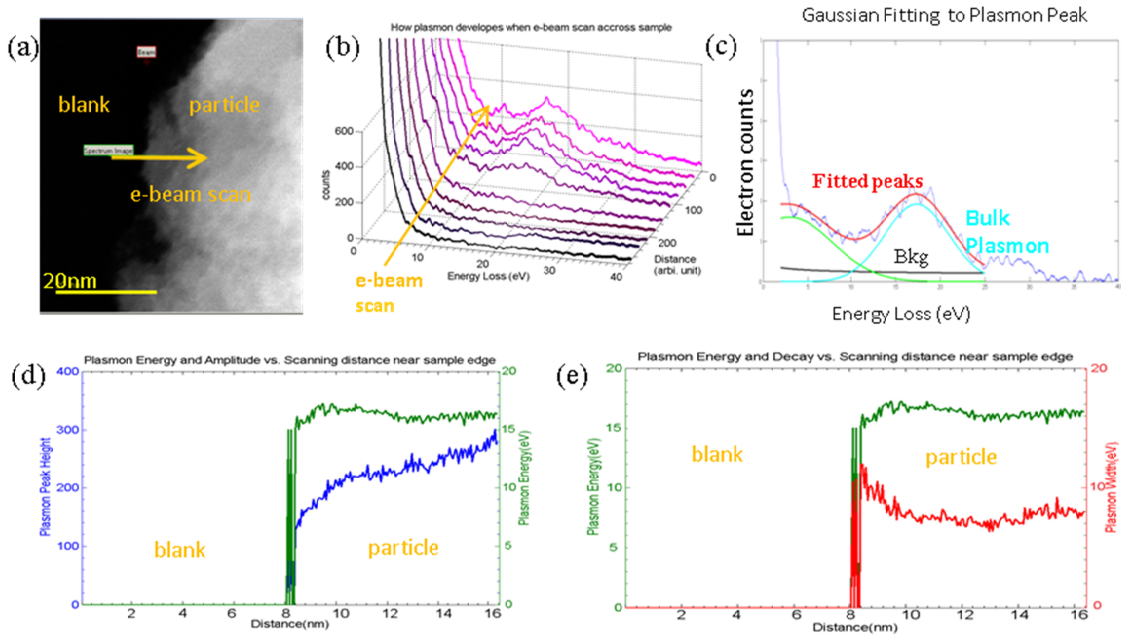

Fig. 6 (a) TEM image and (b) illustration of EELS line scan. When electron beam scans from blank to the particle, plasmon peak is developed. (c) peak fitting to acquire peak height, width and position (energy). (d)(e)fitted result as a function of scanning position. When e-beam comes across the sample(~8nm), the energy positions (green curves) are almost constant, while the amplitude increases when electron-beam goes deeper inside the sample; the width is constant in the bulk region while increased near the interface.

*Fig. 7*

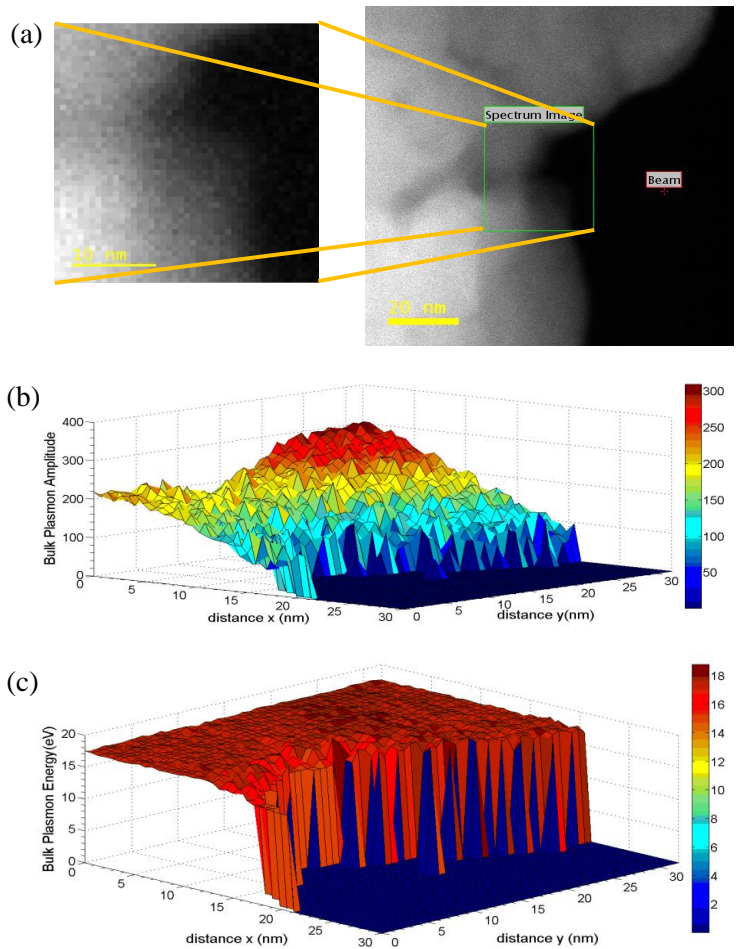

Fig.7. (a) TEM image of the Bi2Se3 particle, and the selected 2D area for EELS spectrum image. Each position in the left 48×45 square corresponds to an EELS spectrum.

(b) The bulk plasmon amplitude as a function of positions. This amplitude is zero in the blank region, and gradually increases inside the sample bulk. This behavior is similar to the result of line-scan.

(c) The bulk plasmon energy is almost independent of position in the sample. However, it is proportional to the square root of the number of carrier density. A little higher value indicate an increase of larger carrier density compared with single crystal.

1  **Supplementary Materials**

2  **S1 First-order correction to topological-induced SP energy change**

3  The dispersion relation is written done in eq. (6). In the non-topological case, LHS=0, the
4  dispersion relation can be written as:

$$\frac{1}{\sqrt{a^2+(q_0^2-w_0^2)}-a}+\frac{\left(1-\frac{1}{w_0^2}\right)}{\sqrt{a^2+(q_0^2-w_0^2+1)}+a}=0 \quad \text{(S1)}$$

6  Since LHS is a small value (<1E4), we assume it will perturb the energy only a little, i.e.
7  at same momentum $q_0$, the energy is shifted a little: the solution becomes ($q_0$, $\omega_0$+d$\omega$),

8  $d\omega \ll \omega_0$. In this case, if we define $A \equiv a^2 + q_0^2 - w_0^2$, we have A>0 since real SP modes
9  lie below the light line and $\omega_0 < q_0$ is always valid. Thus the perturbation of A can be
10  written as

$$\left(\frac{\alpha\theta}{\pi}\right)^2 = -\left(\frac{dA}{2\sqrt{A}}+\frac{dA}{2\sqrt{A+1}}\right)\left(\frac{1}{\sqrt{A}-a}+\frac{1-\frac{1}{\omega_0^2}}{\sqrt{A+1}+a}\right) -$$

$$\left(\sqrt{A+1}+\sqrt{A}\right)\left[\frac{dA}{2\sqrt{A}\left(\sqrt{A}-a\right)^2}-\frac{\left(1-\frac{1}{\omega_0^2}\right)dA}{2\sqrt{A+1}\left(\sqrt{A+1}+a\right)^2}-\frac{dA}{\omega_0^4\left(\sqrt{A+1}+a\right)}\right] \quad \text{(S2)}$$

12  from which the first order perturbation of SP energy $d\omega$ can be obtained.

14  **S2 Eigen equation with topological axion term**

15  From eq (9), and using recursive relations of modified Bessel functions, be aware that the
16  derivative is respect to the imaginary value while the arguments are taken to be real, we
17  have

$$0 = \frac{q_z^2 n^2}{d^2}\left[\frac{\left(1-\frac{1}{\omega^2}\right)-\mu_1}{k_1 k_2}\right]^2 \omega^2 - \alpha^2 \left(\frac{n}{|k_2 d|}+\frac{I_{n+1}(|k_2 d|)}{I_n(|k_2 d|)}\right)\left(\frac{n}{|k_1 d|}-\frac{K_{n+1}(|k_1 d|)}{K_n(|k_1 d|)}\right)\mu_1 k_1 k_2 + \quad \text{(S3)}$$

$$\left[\left(\frac{n}{|k_2 d|}+\frac{I_{n+1}(|k_2 d|)}{I_n(|k_2 d|)}\right)\left(1-\frac{1}{\omega^2}\right)k_1 - \left(\frac{n}{|k_1 d|}-\frac{K_{n+1}(|k_1 d|)}{K_n(|k_1 d|)}\right)k_2\right]\left[\left(\frac{n}{|k_2 d|}+\frac{I_{n+1}(|k_2 d|)}{I_n(|k_2 d|)}\right)k_1 - \mu_1\left(\frac{n}{|k_1 d|}-\frac{K_{n+1}(|k_1 d|)}{K_n(|k_1 d|)}\right)k_2\right]$$

**S3 Energy-Dispersive X-ray Spectroscopy (EDS) of $Bi_2Se_3$ nanoparticles**

EDS of of $Bi_2Se_3$ nanoparticles is shown in Fig. S1. The characteristic peaks of Bi and Se elements are seen clearly. Since neither Carbon nor Copper show low energy-loss EELS peaks near the bulk plasmon energy [42-44], the peak ~17eV can be excluded from the contribution of Cu or Carbon.

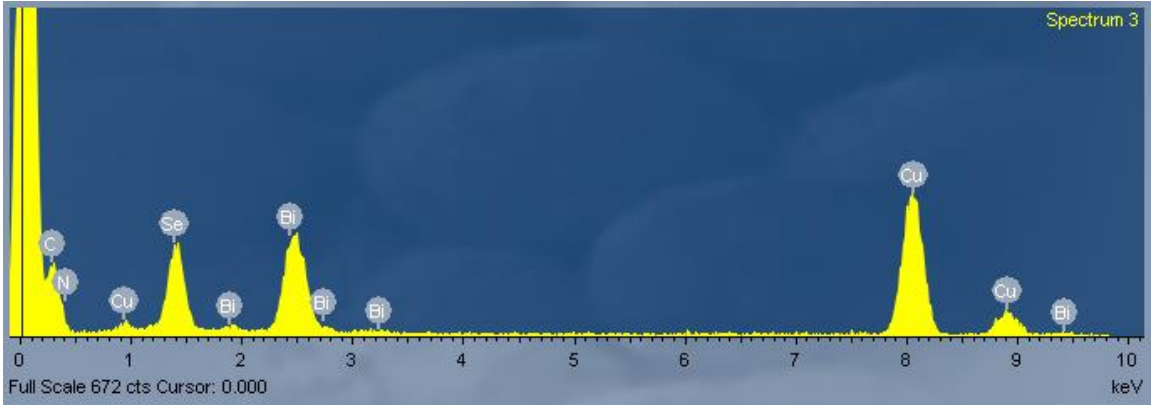

Fig.S1. EDS for Bi2Se3 nanoparticles. The characteristic X-rays of Bi and Se are clearly seen. The Cu and C are caused by the TEM sample grid.

**S4 Topological Effects to plasmon eigenmodes N=0,2,3 at various core radii**

Fig. 5 (c, d) shows the size effect of topological-induced energy shift, which can be used to be compared with the result from the ansatz Fig.2 directly. In order to see the how topological effect is varying in different eigenmodes, the energy shifts are also plotted at fixed values of radii, see in Fig. S2. The N=0 are identical to the discussion of the ansatz, where the energy shift goes down to zero at $q$ tends to zero. However, when N>0, the energy shift shows a non-dispersive feature, which is similar to the non-dispersive relation itself. This is consistent in the framework of perturbation of Maxwell's equation in eq. (10). When N goes higher, this SP energy difference will be pretty much the same, tends to a single curve at high-N limit. What's more, when the core radius becomes larger, the different eigenmodes also collapse into a single curve. As seen in Fig. S2 (f), at d=1000nm, different eigenmodes already overlap with each other and cannot be distinguished anymore.

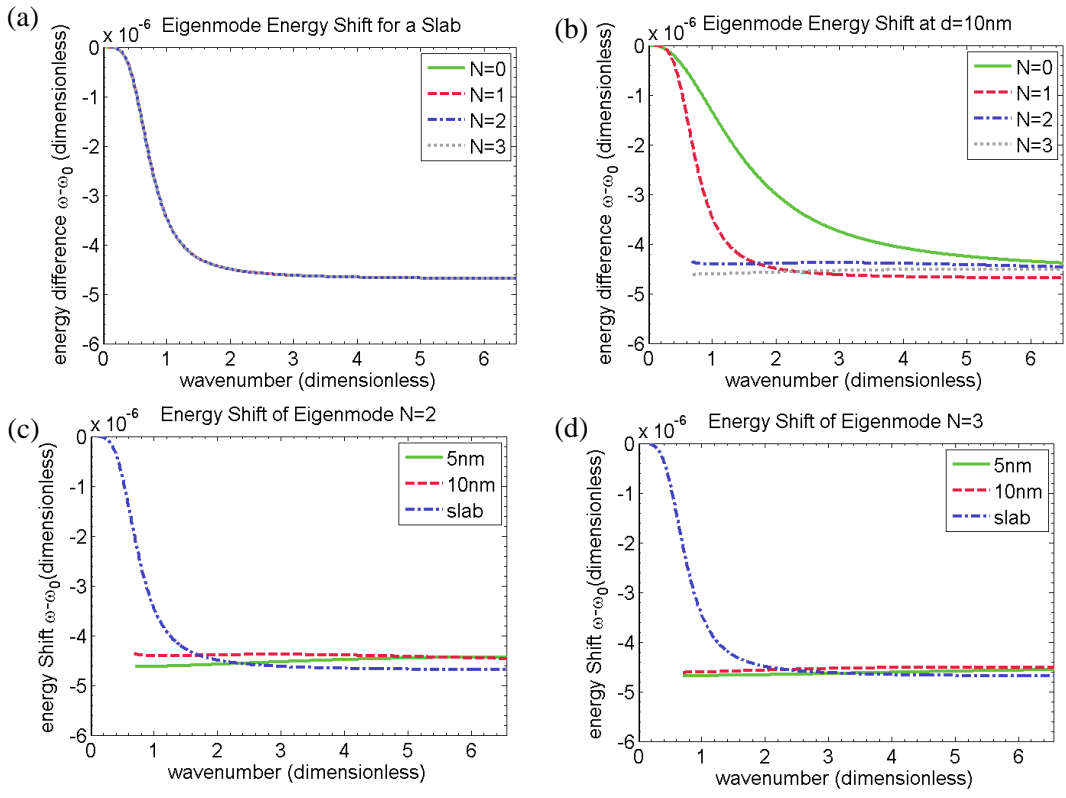

Fig. S2. (a-f) Topological induced energy red-shift $\omega_{topo} - \omega_0$ of cylindrical SP eigenmodes N=0,2,3 at radii 5nm, 10nm and slab, respectively.